\documentstyle[aps,prl,multicol,epsf]{revtex}
\begin{document}
\draft
\title{Quantum-disordered slave-boson theory of 
underdoped cuprates
}

\author{Yong Baek Kim$^a$ and Ziqiang Wang$^b$}
\address{
$^a$ Department of Physics, The Ohio State University,
Columbus, OH 43210, U.S.A. \\
$^b$ Department of Physics, Boston College, Chestnut Hill, 
MA 02167, U.S.A.  
}

\maketitle

\begin{abstract}

We study the stability of the spin gap phase 
in the $U(1)$ slave-boson theory of the $t-J$ 
model in connection to the underdoped cuprates.
We approach the spin gap phase from the superconducting
state and consider the quantum phase transition of the 
slave-bosons at zero temperature by introducing 
vortices in the boson superfluid.
At finite temperatures, the properties
of the bosons
are different from those in the strange metal phase
and lead to modified gauge field fluctuations.
As a result, the spin gap phase can be stabilized
in the quantum critical (QC) and quantum disordered (QD) regime
of the boson system.    
We also show that the regime of QD
bosons with the paired fermions can be regarded
as a strong coupling version of the recently proposed
nodal liquid theory. 

\end{abstract}


\begin{multicols}{2}

The pseudogap behavior of underdoped cuprates in various physical 
properties has been a subject of intensive research 
recently\cite{review}.
The importance of the subject comes from the fact that 
understanding this behavior and relating it to the properties 
of the superconducting state may give us an important clue
for the mechanism of the superconductivity. 
Among many different theories for the pseudogap, one of the
earliest proposals was the mean field theory of the $t-J$ model
which is an effective low energy theory of the Hubbard model in the limit of 
large on-site Coulomb repulsion\cite{mf}.

In order to explain why this mean field theory is appealing,
let us begin with the slave boson representation
of the $t-J$ model. 
In the strong coupling limit, the double occupancy of the
electrons at each site is prohibited, thus
it is convenient to describe the Hilbert space of the
electrons in terms of a neutral spin 1/2 fermion, 
$f^{\dagger}_{i \alpha}$, 
representing the singly occupied sites with a spin up or spin 
down electron, and a spinless charge $e$ boson, $b_i$,  
keeping track of empty sites.
As a result, the electron
operator can be written as $c^{\dagger}_{i \alpha} = 
f^{\dagger}_{i \alpha} b_i$ with the constraint 
$\sum_{\alpha} f^{\dagger}_{i \alpha} f_{i \alpha} + 
b^{\dagger}_i b_i = 1$.
The $t-J$ model can be written as
\begin{eqnarray}
{\cal L} &=& \sum_{i \alpha} f^{\dagger}_{i \alpha} 
(\partial_{\tau} - \mu) f_{i \alpha} 
+ \sum_i b^{\dagger}_i (\partial_{\tau} - iA_0) b_i \cr
&&- t \sum_{\langle ij \rangle, \alpha} 
e^{-iA_{ij}} b_i b^{\dagger}_j f_{i \alpha}^{\dagger} 
f_{j \alpha} - J \sum_{\langle ij \rangle, \alpha \beta}
f^{\dagger}_{j \alpha} f_{i \alpha} f^{\dagger}_{i \beta} 
f_{j \beta} \cr
&&- i \sum_i a_{0i}
( \sum_{\alpha} f^{\dagger}_{i \alpha} f_{i \alpha} + 
b^{\dagger}_i b_i - 1) \ ,
\label{tJ}
\end{eqnarray}
where $A_{ij} = \int^j_i {\bf A} \cdot d{\bf l}$ and $A_0$ 
represent the external vector and scalar potentials.
Here $a_{0i}$ is the Lagrange multiplier enforcing the
local constraint.
Using the Hubbard-Stratonovich transformation, the action can be 
rewritten as
\begin{eqnarray}
{\cal L} &=& \sum_{\langle ij \rangle} \left [ 
J|Q_{ij}|^2 + {|\Delta_{ij}|^2 \over J} \right ] + 
{\cal L}_F + {\cal L}_B \ , \cr
{\cal L}_F &=& \sum_{i \alpha} f^{\dagger}_{i \alpha} 
(\partial_{\tau} - \mu - ia_{0i}) f_{i \alpha} \cr
&&- J \sum_{\langle ij \rangle, \alpha} \left [ 
|Q_{ji}|e^{-ia_{ij}} f^{\dagger}_{i \alpha} f_{j \alpha} + 
{\rm H.c.} \right ] \cr
&&+ \sum_{\langle ij \rangle, \alpha \beta} \left [ \Delta_{ij} 
\epsilon^{\alpha \beta} f_{j \beta} f_{i \alpha}
+ {\rm H.c.} \right ] \ , \cr
{\cal L}_B &=& \sum_i b^{\dagger}_i 
(\partial_{\tau} - ia_{0i} - iA_0) b_i \cr
&&- t \sum_{\langle ij \rangle} 
\left [ |Q_{ji}|e^{-ia_{ij}} e^{-iA_{ij}}
b^{\dagger}_i b_j + {\rm H.c.} 
\right ] \ ,  
\label{actionfb} 
\end{eqnarray}
where $Q_{ij} = |Q_{ij}|e^{ia_{ij}}$.
In the mean field theory, it was found that $|Q_{ij}|=Q$ = const.
and $a_{ij}=0$. 
It has been established that there are four different phases at the 
mean field level depending on the values of $\Delta_{ij}$ and 
$\langle b_i \rangle$\cite{mf}.
i) Superconducting phase:
$\Delta_{ij} = \langle \epsilon^{\alpha \beta} 
f^{\dagger}_{i \alpha} f^{\dagger}_{j \beta} 
\rangle \not= 0$ and
$\langle b_i \rangle \not= 0$.
ii) Spin gap phase:
$\Delta_{ij} \not= 0$ and
$\langle b_i \rangle = 0$.
iii) `Fermi liquid' phase:
$\Delta_{ij} = 0$ and
$\langle b_i \rangle \not= 0$.
iv) `Strange metal' phase:
$\Delta_{ij} = 0$ and
$\langle b_i \rangle = 0$.
As the electron is a combination of the fermion
and boson, an excitation gap of the electron will be
generated from the gap of the spin carrying fermions in the spin gap
phase.
Therefore, this theory already suggests the spin gap phase as
a possible candidate for the pseudogap behavior of underdoped cuprates.   

However, it was later found by Ubbens and Lee that 
the spin gap phase of the mean field theory is unstable 
against the fluctuations about the mean field state when
the spin gap phase is approached from the 
strange metal phase\cite{ubbens}. 
One can see from Eq.\ref{actionfb} that the 
fluctuation in the phase of $Q_{ij}$ can be represented
in terms of a $U(1)$ gauge field {\bf a} through 
$a_{ij} = \int^j_i {\bf a} \cdot d{\bf l}$. 
This is associated with the internal $U(1)$ symmetry in the 
slave-boson representation of the $t-J$ model.
The previous study mentioned above would imply that 
the $U(1)$ slave-boson theory does not support the spin 
gap phase. 

In order to make the later discussion
more concrete, let us first reproduce the arguments
of Ubbens and Lee \cite{ubbens}.
Upon approaching the spin gap phase from the
strange metal phase, one can
evaluate the free energy cost for opening up
the gap for the fermions in the following fashion.
First we evaluate the contribution to the 
free energy from the gauge field fluctuations
in the strange metal phase.
\begin{equation}
F_g = \sum_{\bf q} \int {d\omega \over 2 \pi}
(2 n_B (\omega) + 1) {\rm arctan} 
\left ( {{\rm Im}D^{-1} \over 
{\rm Re}D^{-1}} \right ) \ ,
\label{freeenergy}
\end{equation}
where $D ({\bf q},\omega)$ is the gauge field
propagator and given by 
$D^{-1} ({\bf q},\omega) = 
\Pi^{jj}_F ({\bf q},\omega) + \Pi^{jj}_B ({\bf q},\omega)$.
Here $\Pi^{jj}_F$ and $\Pi^{jj}_B$ are the fermion and
boson current-current correlation functions
respectively. 
In the strange metal phase, $\Pi^{jj}_F$ and $\Pi^{jj}_B$ were
assumed to have the free fermion and boson forms:
$\Pi^{jj}_F  = -i \gamma \omega / q + \chi_F q^2$
and $\Pi^{jj}_B \approx \chi_B q^2$.
Here $\gamma = 2n_e/k_F$, $\chi_F = 1/(12 \pi m_F)$, and 
$\chi_B = (e^{T_{BE}/T}-1)/(24 \pi m_B)$, where 
$n_e$ is the electron density, $m_F = 1/(2JQ)$, 
$m_B = 1/(2tQ)$, and $T_{BE} = 2 \pi x / m_B$ ($x$ is
the doping concentration) is the mean 
field boson condensation temperature.
As a result, $D^{-1}$ is given by
\begin{equation}
D^{-1} \approx -i \gamma {\omega \over q} + \chi q^2 \ ,
\label{gauge}
\end{equation} 
where
$\chi = \chi_F + \chi_B$.
Using Eq.\ref{freeenergy} and Eq.\ref{gauge}, 
we get $F_g \propto T^{5/3}$.
When the gap is opened, $\Delta$ can be introduced to 
cutoff the frequency integral and the effect is simply
replacing $T$ by $\Delta$ in $F_g$. Thus, the free
energy cost for opening up the spin gap is proportional to
$\Delta^{5/3}$. Since the mean field pairing energy
gain is proportional to $-\Delta^2$, the free energy cost
from the gauge field always dominates. Thus the spin gap
phase cannot be stabilized.  
In their work, the bosons
in the strange metal phase were 
assumed to behave rather classically. 
The transition from the strange metal phase to the 
spin gap phase occurs due to the pairing of the 
fermions while the bosons were assumed to be 
still classical.  

In this paper, we suggest that the spin gap phase 
can be stabilized at low temperatures because the 
properties of the bosons are {\it different} from 
those in the high temperature strange metal phase. 
Motivated by the phase diagram of the cuprates, we 
suggest that at zero temperature there exists a quantum 
disordering phase transition of the bosons driven by vortices  while 
the fermions remain paired across the transition (Fig.~1). 
The low density boson system ($x < x_c$) becomes
an insulator due to the condensation of vortices.
On the other hand, if $x > x_c$, the bosons are 
in the superfluid state while the vortices are in 
the insulating state. 
Recall that the electron resistivity
is given by $\rho_e = \rho_F + \rho_B$, where $\rho_F$
and $\rho_B$ are the fermion and boson resistivities 
respectively.
Since $\rho_F = 0$ and $\rho_B$ diverges when $x < x_c$, 
the zero temperature spin gap phase is an insulator.  
We found that, near the QC point of the 
phase transition, the properties of the
bosons are significantly modified. 
Using the boson-vortex duality, we obtain the 
current-current correlation function, $\Pi^{jj}_B$, 
for the QD bosons. 
Due to the modified bosonic properties,
the gauge field propagator has a different form at low temperatures,
which leads to, according to Eq.\ref{freeenergy}, the free
energy cost for opening the spin gap smaller than the pairing
energy gain, $-\Delta^2$. Thus the spin gap phase can be
stabilized. 
In addition, we suggest that the recently proposed 
nodal liquid theory can be regarded as a weak coupling
version of our theory if the constraints are 
ignored\cite{nordal}.

Let us begin with the superconducting state where
the fermions are paired ($\Delta_{ij} \not= 0$) and 
the bosons are in the superfluid state 
($\langle b_i \rangle \not= 0$).
The phase diagram of the cuprates tells us that the
superconducting state exists when the doping 
concentration is larger than a particular value $x_c$.
The question is whether the normal state in the regime 
$x < x_c$ can be obtained by suppressing the boson
superfluid while the fermions are still paired.
In order to answer this question, we have to understand 
the nature of the quantum disordering phase transition of the bosons
across $x_c$. 
We suggest that the latter is
due to the condensation of vortices.
The vortices carry the flux quantum $hc/e$ because
the bosons carry charge $e$.
In order to describe the QD state of
bosons, it is convenient to use the dual representation
of the bosons\cite{dual}. 
Notice that the external electromagnetic fields only
couple to the bosons. 
The continuum limit of
${\cal L}_B$ in the presence of the external fields is given by
\begin{equation}
{\cal L}_B = b^{\dagger} 
(\partial_{\tau} - ia_0 - iA_0) b
- {1 \over 2m_B} b^{\dagger} (\nabla - i{\bf a} 
- i{\bf A})^2 b \ .  
\label{boson}
\end{equation}
Following Ref.\cite{dual}, 
let $b = \sqrt{\rho} \phi$ where $\rho$ is
positive definite, which corresponds to the boson density, and 
$\phi$ is a unimodular complex field satisfying
$\phi^{\dagger} \phi = 1$.
Then the action (Eq.\ref{boson}) becomes
\begin{eqnarray}
{\cal L}_B &=& i \rho (\phi^{\dagger} 
{\partial_{\tau} \over i} \phi - a_0 - A_0) \cr
&&+ {\rho \over 2m_B} \left | \phi^{\dagger} 
{\nabla \over i} \phi - {\bf a} - {\bf A} \right |^2 
+ {1 \over 2m_B} \left |\nabla \sqrt{\rho} \right |^2 \ .    
\end{eqnarray}
Next, we decouple the second term by a Hubbard-Stratonovich
transformation, which leads to
\begin{eqnarray}
{\cal L}_B &=& i J_{\mu} 
(\phi^{\dagger} {\partial_{\mu} \over i} \phi - a_{\mu} - A_{\mu}) \cr
&&+ {m_B \over 2 \rho} \left | {\bf J} \right |^2 +
{1 \over 2m_B} \left | \nabla \sqrt{\rho} \right |^2 \ ,
\end{eqnarray}
where $J_{\mu} = (\rho,{\bf J})$ is the boson three-current and
$A_{\mu} = (A_0,{\bf A})$.
In order to isolate the vortices, we write $\phi$ as
$\phi = \phi_v e^{i \theta}$,
where $\theta$ is single valued and $\phi_v$ represents
the vortices.
Integration over $\theta$ gives the continuity constraint
$\partial_{\mu} J_{\mu} = 0$.
One can solve this constraint by introducing a new field
$M_{\lambda}$ through $J_{\mu} = \epsilon_{\mu \nu \lambda} 
\partial_{\nu} M_{\lambda} \equiv (\partial \times M)_{\mu}$. 
It is also useful to introduce $\delta M_{\mu} = M_{\mu} - 
{\bar M}_{\mu}$ with 
$(\partial \times {\bar M})_0 = x$ and 
$(\partial \times {\bar M})_{1,2} = 0$, where $x$ is 
the average boson density.
Then the action in the long wavelength limit can be 
written as
\begin{eqnarray}
{\cal L}_B &=& {m_B \over 2 x} 
(\left | (\partial \times \delta M)_1 \right |^2 + 
\left | (\partial \times \delta M)_2 \right |^2) \cr
&&+ i \ \delta M_{\mu} [J^v_{\mu} - 
(\partial \times a)_{\mu} 
- (\partial \times A)_{\mu} ] \ ,
\end{eqnarray}
where $(\partial \times a)_{\mu} \equiv 
\epsilon_{\mu \nu \lambda} \partial_{\nu} a_{\lambda}$
and $J^v_{\mu}$ is the vortex three-current
$J^v_{\mu} = \epsilon_{\mu \nu \lambda} \partial_{\nu}
\phi^{\dagger}_v {\partial_{\lambda} \over i} \phi_v
\equiv (\partial \times \phi^{\dagger}_v {\partial \over i}
\phi_v)_{\mu}$.
When the vortices are condensed, the vortex transverse 
current-current correlation function in the long wavelength and
low energy limit should have the form:
$\langle J^v J^v \rangle = C \rho^s_v (T,x)$, where 
$\rho^s_v (T,x)$ is the superfluid density of the vortex 
condensate and $C$ is a constant. 
Using this correlation function and integrating out $M_{\mu}$
degrees of freedom, the effective action for the transverse part 
of $a_{\mu} + A_{\mu}$ field becomes 
\begin{equation}
{\cal L}_{B,{\rm eff}} = \sum_{{\bf q},\omega}
{q^2 \over C\rho^s_v (T,x)} 
|({\bf a} + {\bf A})_{{\bf q},\omega}|^2 \ . 
\end{equation}
Comparing with Eq.\ref{boson}, we can read off the boson
current-current correlation function at finite 
temperatures:
\begin{equation}
\Pi^{jj}_B = {q^2 \over C\rho^s_v (T,x)} \ . 
\end{equation}
Since $\rho^s_v$ becomes smaller and smaller as
the critical point is approached, if we write down
$\Pi^{jj}_B$ as $\Pi^{jj}_B = \chi_B q^2$ with 
$\chi_B \propto 1/C\rho^s_v$, we have 
$\chi_F \ll \chi_B$ near the critical point. 
Then the gauge field propagator at finite temperatures 
can be written as
\begin{equation}
D^{-1} \approx - i \gamma \omega / q + 
{q^2 \over C\rho^s_v} \ .
\end{equation}
Using the above propagator and Eq.\ref{freeenergy}, 
we obtain
\begin{equation}
F_g \propto [\rho^s_v (T,x)]^{2/3} T^{5/3} \ .
\label{Fgauge}
\end{equation}
Now some remarks on the temperature dependence of
$\rho^s_v (T,x)$ are in order.
Using dimensional analysis of hyperscaling, {\it i.e.} the free energy
within a correlation volume $\xi^d\xi^z$ is non-singular in the vicinity
of the quantum critical point, the scaling form of 
$\rho^s_v (T,x)$ in $d$-dimensions can be written as \cite{fgg}
\begin{equation}
\rho^s_v (T,x) = \xi^{2-d-z} F(\xi^{z} T) \ ,
\end{equation}
where $\xi \propto |x-x_c|^{-\nu}$ is the correlation
length of the vortex superfluid and $F(x)$ is 
a scaling function.
Here $z$ is the dynamical critical exponent and 
$\nu$ is the correlation length exponent.
If $\xi^z T \gg 1$, {\it i.e.}in the quantum critical regime,
the vortex superfluid density must be independent of $\xi$
and only depend on temperature. In two dimensions, this 
implies $F(x) \sim x$ for $x \gg 1$, leading to $\rho^s_v \propto T$.
Substituting the latter into Eq.\ref{Fgauge}, we find that the
free energy cost for opening up the spin gap due to the gauge field
becomes $F_g \sim T^{7/3}$ in the QC regime.
Similarly, in the quantum disordered regime,
$\xi^z T\ll1$, the scaling function $F(x)\to {\rm const}$ for $x\ll1$,
$\rho_v^s$ becomes temperature independent $\rho_v^s\propto\xi^{-z}$.
In this case, Eq.\ref{Fgauge} becomes $F_g\propto \xi^{-2z/3}T^{5/3}$.
When $T < \Delta$, one would use $\Delta$ as a low frequency 
cutoff instead of $T$. As long as $\Delta > B |x-x_c|^{z\nu}$,
one can trade $T$ and $\xi^{-z}$ with $\Delta$ in the above expression
to obtain the energy cost for opening up the gap 
$F_g\propto\Delta^{7/3}$.    
Thus the mean field fermion pairing energy gain, $-\Delta^2$, 
wins and the spin gap phase can be stabilized in both the QC and the QD
regimes.
We stress that this is a {\it qualitatively} very different
behavior compared to the previous claim\cite{ubbens}.

\begin{figure}    
\vspace{-0.5truecm}
\center    
\centerline{\epsfysize=2.2in    
\epsfbox{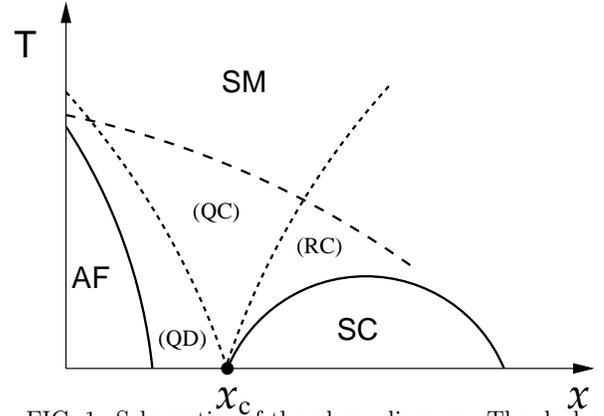}}
\vspace{-0.5truecm}    
\begin{minipage}[t]{8.1cm}    
\caption{Schematics of the phase diagram.
The dashed lines indicate the crossover between QC and QD,
QC and RC, and QC/RC and the strange metal (SM) regimes. 
The spin gap is stable in the QC and QD regimes.}  
\label{fig2}   
\end{minipage}    
\end{figure}    

At sufficiently high temperatures, the bosons would behave
classically. In this regime, the argument of Ubbens 
and Lee can be applied\cite{ubbens}.
Therefore, there should be a crossover from either QD or 
QC regime to the high temperature
strange metal phase where the gauge field fluctuations destroy 
the spin gap phase.
Fig.~1 shows a schematic phase diagram in the $T-x$ plane. 
Notice that there exists a crossover line between the 
strange metal phase and the spin gap phase
stabilized in both the QD or QC regime of bosons. 
The dashed lines ($T = B|x-x_c|^{z\nu}$ line) represent the 
crossovers between the QD, QC, and 
renormalized classical (RC) regimes of bosons. 
Since the bosons in the RC 
regime behave essentially in a classical fashion,
the spin gap phase is not stable there.
Thus, the spin gap phase is restricted to the 
underdoped regime.
The crossover from the QC regime to RC
regime occurs when $T < B (x-x_c)^{z\nu}$ and $x > x_c$.
Thus the transition from the strange metal phase to the
superconducting phase is likely to occur in the RC regime 
({\it i.e.}, $T_c$ may be lower than $B (x-x_c)^{z\nu}$) 
so that it is characterized by a more or less
classical transition of the bosons to the superfluid 
state.  
On the other hand, if $T < B (x_c-x)^{z\nu}$ and $x < x_c$, 
the crossover from the QC regime to the 
QD state of bosons occurs.
Thus the antiferromagnetic phase is in the quantum
disordered regime of the bosons.    
In the low temperature limit, the relevant excitations
in d-wave spin gap phase would be the neutral Dirac 
fermions. At lower doping concentrations, the $SU(2)$
fluctuations \cite{donhkim} may become important and 
these neutral Dirac 
fermions may be confined when $x$ is
sufficiently small and the antiferromagnetism 
occurs \cite{donhkim,affleck}. 

Now let us discuss the relation between the present
theory and the nodal liquid theory proposed recently,
where the quantum disordering of Cooper pairs was
discussed \cite{nordal}.
Let us begin with
the continuum limit of ${\cal L}_F$ assuming that 
$\Delta_{ij}=\Delta$ is 
a uniform complex number and we will comment on the
d-wave case later:
\begin{eqnarray}
{\cal L}_F &=& f^{\dagger}_{\alpha}  
(\partial_{\tau} - \mu - ia _0) f_{\alpha} 
- {1 \over 2m_F} f^{\dagger}_{\alpha} 
(\nabla - i{\bf a})^2 f_{\alpha} \cr
&&+ \Delta \epsilon^{\alpha \beta} f_{\beta} f_{\alpha}
+ \Delta^* \epsilon^{\alpha \beta} f^{\dagger}_{\alpha} 
f^{\dagger}_{\beta} \ .  
\end{eqnarray}
Let $\Delta = |\Delta|e^{i\phi}$. After doing the gauge 
transformation
$f_{\alpha} \rightarrow f_{\alpha} e^{i \phi/2}$ and
$b \rightarrow b e^{i \phi/2}$, we obtain the action
in the absence of the external fields:
\begin{eqnarray}
{\cal L}_F &=& f^{\dagger}_{\alpha} 
(\partial_{\tau} - \mu - i{\tilde a}_0) f_{\alpha} 
- {1 \over 2m_F} f^{\dagger}_{\alpha} 
(\nabla - i{\tilde {\bf a}})^2 f_{\alpha} \cr
&&+ |\Delta| (\epsilon^{\alpha \beta} f_{\beta} f_{\alpha} 
+ \epsilon^{\alpha \beta} f^{\dagger}_{\alpha} 
f^{\dagger}_{\beta}) \ , \cr
{\cal L}_B &=& b^{\dagger} 
(\partial_{\tau} - i{\tilde a}_0) b
- {1 \over 2m_B} b^{\dagger} 
(\nabla - i{\tilde {\bf a}})^2 b \ , 
\end{eqnarray}  
where ${\tilde a_{\mu}} = a_{\mu} + (\partial_{\mu} \phi)/2$.
This amounts to fixing the gauge for $a_{\mu}$.
In the superconducting state, the order parameter in the strong coupling
limit can be written as 
$\Delta^{e} = \langle \epsilon^{\alpha \beta}
c^{\dagger}_{\alpha} c^{\dagger}_{\beta} \rangle
\approx  \langle \epsilon^{\alpha \beta}
f^{\dagger}_{\alpha} f^{\dagger}_{\beta} \rangle
\langle b b \rangle$.
In the gauge choice we have taken, $\langle \epsilon^{\alpha \beta}
f^{\dagger}_{\alpha} f^{\dagger}_{\beta} \rangle$ is always
real. Therefore, in the superconducting state, we get
$\Delta^{e} \approx |\Delta| \langle b b \rangle
\approx |\Delta| b^2_0 e^{i \theta}$.
Here $b = b_0 e^{i \theta/2}$ is taken.
Notice that the phase of $\Delta$ is dictated by the
phase of the bosons and $|\Delta^{e}| = |\Delta| b_0^2$.
In the superconducting state, we substitute $b$ by 
$b_0 e^{i \theta/2}$ in the action. 
Integrating out
${\tilde a}_0, {\bf a}$, we obtain
\begin{eqnarray}
{\cal L} &=& {\tilde {\cal L}}_{F} + 
{\cal L}_{\theta} + {\cal L}_{F,\theta} \ , \cr
{\tilde {\cal L}}_{F} &=& f^{\dagger}_{\alpha} 
(\partial_{\tau} - \mu - {1 \over 2m_F} \nabla^2) 
f_{\alpha} \cr
&&+ |\Delta| (\epsilon^{\alpha \beta} f_{\beta} f_{\alpha} 
+ \epsilon^{\alpha \beta} f^{\dagger}_{\alpha} 
f^{\dagger}_{\beta}) \cr
&&+ \int d^3 r' J^F_{\mu} (r) (\Pi^F)^{-1}_{\mu \nu} (r-r') 
J^F_{\nu}(r') \ , \cr
{\cal L}_{\theta} &=& \kappa_{\mu} 
(\partial_{\mu} \theta)^2 \ , \cr
{\cal L}_{F,\theta} &=& g_{\mu} J^F_{\mu} 
(\partial_{\mu} \theta) \ ,
\label{phase}
\end{eqnarray}
where $J^F_{\mu} = (\rho^F, {\bf J}^F)$ with 
$\rho^F = f^{\dagger}_{\alpha} f_{\alpha}$ and
${\bf J}^F = {1 \over 2 i m_F} (f^{\dagger}_{\alpha} \nabla 
f_{\alpha} - {\rm H.c.})$.
$\Pi^F_{\mu \nu} = \langle J^F_{\mu} J^F_{\nu} \rangle$
is the fermion three-current correlation function.
Here $g_0 = 1/2$ and 
$g_{1,2} = ({x \over m_B})/[{x \over m_B} 
+ {(1-x) \over m_F}]$.
Also $\kappa_0 = N(0)^{-1} = {2\pi \over m_F}$ 
and $\kappa_{1,2} = {1 \over 2}({x \over m_B}) g_{1,2}$.
The continuum action derived above can be also obtained
directly from the lattice model given by 
Eq.\ref{tJ}\cite{wang}.

If $\Delta_{ij}$ has the d-wave symmetry,
the relevant degrees of freedom in the low energy limit 
are the excitations near the nodes.
These excitations have the Dirac spectra
and can be represented by 
$d_a = (d_{a \uparrow}, d^{\dagger}_{a \downarrow})$,
where $a = 1, 2$ represent two pairs of nodes.
Following Ref.\cite{nordal}, the fermion part of the action
becomes
\begin{eqnarray}
{\cal L}_d &=& 
d^{\dagger}_1 (\partial_{\tau} - v_F \tau^z i \partial_x
- v_{\Delta} \tau^x i \partial_y ) d_1 \cr
&&+ d^{\dagger}_2 (\partial_{\tau} - v_{\Delta} \tau^z 
i \partial_y - v_F \tau^x i \partial_x ) d_2 \ ,
\label{d1}
\end{eqnarray}  
where $\tau^{x,y,z}$ are the Pauli matrices and
$v_{\Delta} = \Delta/\sqrt{2}$.
The coupling between the fermion and the phase fields
is also changed to
${\cal L}_{d, \theta} = {\tilde g}_{\mu} 
J^d_{\mu} (\partial_{\mu} \theta)$,
where $J^d_{\mu}$ is the current of the Dirac 
fermion $d$. In this case,
the corresponding action in Eq.\ref{phase} looks similar to 
that of the nodal liquid theory
if the additional terms coming from the constraints 
are ignored. 
Thus we conclude that our theory can be
also interpreted as the strong coupling version of 
the nodal liquid theory. However, due to the presence of
the constraints, the structure of the theory is not exactly the same. 
Notice also that the presence of the
$hc/e$ vortices as well as the $hc/2e$ vortices
was pointed out by Sachdev\cite{sachdev} 
as a consequence of
the constraint imposed by the gauge field.
We believe that the presence of the $hc/e$ vortices
suggested by the nodal liquid theory comes out 
naturally in the present strong coupling theory.
It was also suggested that the vortex condensate supports
charge $e$ solitons (dual vortex of the vortex condensate), 
{\it holons}\cite{nordal} in the dual picture. 
Using the boson-vortex duality, one can see
that the slave-bosons in the boson picture may be the natural 
candidates for the solitonic excitations in the dual 
representation. 

In summary, using the $U(1)$ slave-boson representation of the 
$t-J$ model, we obtained the spin gap phase by quantum disordering 
the slave-boson superfluid of the superconducting phase while
the spin-carrying neutral fermions are paired. 
We found that the spin gap phase can be stabilized
in the QC and QD regime of the bosons against the
fluctuations about the mean field state.    
We also showed that the spin gap phase at zero temperature 
obtained in this fashion can be regarded
as a strong coupling version of the nodal liquid phase. 

We thank M. P. A. Fisher, Hae-Young Kee, G. Vignale, 
and G. Zimanyi for helpful discussions. This work was
initiated when the authors were attending a workshop 
at ITP, UCSB. 
The research at ITP was supported in part by 
NSF grant No. PHY9407194.
We were also supported by the 
A. P. Sloan Foundation (Y.B.K.),
NSF CAREER award DMR-9983783 (Y.B.K.),
and Research Corporation (Z.W.).

\end{multicols}

\end{document}